\documentclass[twocolumn,pra,aps,twocolumn,superscriptaddress]{revtex4-1}
\usepackage{mathtools}
\usepackage{microtype}
\usepackage{epsfig,color}
\usepackage[dvipsnames]{xcolor}

\usepackage[colorlinks=true,citecolor=blue,linkcolor=blue,urlcolor=blue]{hyperref}%
\usepackage{cleveref}

\usepackage{libertine}

\usepackage{dsfont}
\usepackage{todonotes}
\usepackage{easyReview}

\usepackage{physics}

\newcommand{\on}[1]{\operatorname{#1}}
\newcommand{\bs}[1]{\boldsymbol{#1}}

\usepackage{amsmath}
\usepackage{graphicx}
\usepackage{amssymb}

\begin{document}
\title{Ultrafast critical ground state preparation via bang-bang protocols}
\author{Luca Innocenti}
\affiliation{Department of Optics, Palack{\'y} University, 17. Listopadu 12, 771 46 Olomouc, Czech Republic}
\affiliation{Centre for Theoretical Atomic, Molecular and Optical Physics, Queen's University Belfast, Belfast BT7 1NN, United Kingdom}
\author{Gabriele De Chiara}
\affiliation{Centre for Theoretical Atomic, Molecular and Optical Physics, Queen's University Belfast, Belfast BT7 1NN, United Kingdom}
\author{Mauro Paternostro}
\affiliation{Centre for Theoretical Atomic, Molecular and Optical Physics, Queen's University Belfast, Belfast BT7 1NN, United Kingdom}
\author{Ricardo Puebla}
\affiliation{Centre for Theoretical Atomic, Molecular and Optical Physics, Queen's University Belfast, Belfast BT7 1NN, United Kingdom}

\begin{abstract}
The fast and faithful preparation of the ground state of quantum systems is a challenging task but crucial for several applications in the realm of quantum-based technologies. Decoherence poses a limit to the maximum time-window allowed to an experiment to faithfully achieve such desired states. This is of particular significance in critical systems, where the vanishing energy gap challenges an adiabatic ground state  preparation.  We show that a bang-bang protocol, consisting of a time evolution under two different values of an externally tunable parameter, allows for a high-fidelity ground state preparation in evolution times no longer than those required by the application of standard optimal control techniques, such as the chopped-random basis quantum optimization. In addition, owing to their reduced number of variables, such bang-bang protocols are very well suited to optimization purposes, reducing the increasing computational cost of other optimal control protocols. We benchmark the performance of such approach through two paradigmatic models, namely the Landau-Zener and the Lipkin-Meshkov-Glick model. 
Remarkably, the critical ground state of the latter model can be prepared with a high fidelity in a total evolution time that scales slower than the inverse of the vanishing energy gap. 
  \end{abstract}

\maketitle

\section{Introduction}

Quantum technologies have seen considerable progress in recent years~\cite{Dowling:03}, thanks to the unprecedented degree of isolation and manipulation capabilities achieved over individual quantum systems\cite{Leibfried:03,Haroche,Blais:20}, paving the way to the development of novel technologies and furthering our fundamental understanding of quantum information processing~\cite{Nielsen}.
Yet, continued development of these technologies requires fast and robust schemes to prepare and manipulate quantum states. In particular, reducing the preparation time of target quantum states would have a profound impact for several quantum technologies, embodying an area of active research~\cite{Dowling:03,Acin:18}.

The ability to prepare ground states of a given Hamiltonian is especially important for many reasons.
On one hand, arbitrary states can be encoded as ground states of suitably arranged Hamiltonians, which is important for adiabatic quantum computation~\cite{Albash:18}.
On the other hand, the ground state of quantum many-body systems is pivotal to the investigation of quantum phase transitions (QPTs)~\cite{Sachdev}.
Indeed, close to the critical point of a second-order QPT, the ground states feature non-analytic behavior, and are very sensitive to variations of the underlying control parameter.
This provides advantages for tasks such as quantum metrology~\cite{Zanardi:08,Frerot:18,Garbe:20}.
Critical ground states of many-body systems also often possess a large degree of entanglement, making them an invaluable resource for several quantum information tasks~\cite{Osterloh:02,Osborne:02,Vidal:03,Korepin:04,Amico:08,De_Chiara:18}.
Nevertheless, the preparation of a \emph{critical ground state} is experimentally challenging.
This stems from the extremely long time required by adiabatic ground state preparation, due to the vanishing energy gap close to the critical point of a second-order QPT~\cite{Sachdev}.
Devising fast and robust protocols for the generation of critical ground states
is thus an important avenue of research. Such efforts would shed further insight into the study of QPTs, such as the experimental determination of their universality class and the fundamental time constraints posed by their vanishing energy gap. 
Here, we will focus on the preparation of the ground state of a second-order quantum critical model, aiming to shorten the time duration of the protocol.

Currently known fast state-preparation strategies include local adiabatic protocols~\cite{Roland:02,Richerme:13,Quan:10}, shortcuts to adiabaticity~\cite{Torrontegui:13,Odelin:19,Abah:20} and fast quasi-adiabatic ramps~\cite{MartinezGaraot:15}.
These methods typically require the system to be analytically solvable or numerically treatable. In addition, further demanding control in the system, embodied for instance by additional time-dependent parameters, is often required.
Quantum optimal control (QOC), on the other hand, is preferable because of its wide applicability~\cite{glaser2015training,rembold2020introduction,wilhelm2020introduction}.
Such approach has proven valuable in a variety of contexts, including the research for optimal NMR pulses~\cite{nielsen2010optimal,sorensen2018quantum,sorensen2020optimization,batra2020pushpull}.
Common QOC techniques include the Krotov method~\cite{krotov1995global,maximov2008optimal,Goerz:19}, gradient-ascent pulse engineering (GRAPE)~\cite{Khaneja:05,goodwin2016modified,defouquieres2011second,dalgaard2020hessianbased}, chopped-random basis quantum optimization (CRAB)~\cite{Doria:11,Caneva:11,Rach:15,vanFrank:16}, and machine-learning-based approaches~\cite{Bukov:18,Yang:18,Zhang:18,dalgaard2020global,an2019deep,niu2019universal}.
The associated numerical tasks are rarely solvable exactly, and generally require non-trivial numerical optimization techniques. This is further highlighted by studies analyzing the complexity of control landscapes of many-body systems~\cite{Kallush:14}.

In this paper, we show that even simple protocols can provide remarkable results, in some cases even outperforming algorithms as sophisticated as CRAB. 
We showcase this, in particular, for the task of ground state preparation close to a second-order QPT.
We propose the use of a double-bang protocol, which consists of two constant evolutions under a Hamiltonian with fixed parameters rather than a single one as considered in~\cite{Balasubramanian:18,Cohn:18}.
We focus on two paradigmatic models: the Landau-Zener (LZ)~\cite{Zener:32}, and the Lipkin-Meshkov-Glick (LMG) one~\cite{Lipkin:65}.
The latter describes an interacting quantum many-body system featuring a mean-field second-order QPT~\cite{Dusuel:04,Leyvraz:05,Vidal:07,Ribeiro:07,Ribeiro:08}.
Furthermore, we provide strong numerical evidence in support of the optimality of double-bang protocols. Remarkably, our approaches are computationally resource-efficient owing to the small number of parameters defining the protocol.
At the same time, they allow us to reach almost-unit fidelities in  quite short times, compared to pulse shapes obtained via state-of-the-art QOC methods such as CRAB.
As further evidence of the good performances of bang-bang protocols, we show that the time required to achieve the critical ground state of the LMG model with good fidelity scales slower than the inverse of the minimum energy gap, which is the type of scaling observed in previous analyses~\cite{Caneva:11,Caneva:11a}.

The remainder of this paper is organised as follows. In Sec.~\ref{sec:QSL} we formulate the problem while in Sec.~\ref{sec:OC} we discuss the application of optimal control techniques to the ground state preparation problem, focusing on bang-bang protocols. In Sec.~\ref{sec:applications} we showcase the performance and advantages of our method through its application to the LZ and LMG models. Finally, in Sec.~\ref{sec:conclusions} we summarise our main findings and briefly discuss further avenues of investigation.

\section{Ground state preparation and fundamental quantum limits}
\label{sec:QSL}

Let us consider a Hamiltonian $H$ which, without loss of generality, we can assume to depend on a single tunable and dimensionless parameter $g$ according to the decomposition 
\begin{equation}
\label{eq:Ht}
	H(g)=H_0+g H_1,
\end{equation}
where $H_{0,1}$ are time-independent Hamiltonian operators. 
Given the initial and final values $g_0$ and $g_1$ of the (externally controllable) parameter, our goal is to find a time-dependent protocol $g(t)$ such that $\ket{\phi_0(g_0)}$ evolves into $\ket{\phi_0(g_1)}$ in the shortest possible evolution time $\tau$, where $\ket{\phi_0(g)}$ denotes the ground state of $H(g)$.
In general, the associated dynamics cannot be solved exactly, making it necessary to resort to numerical optimization techniques.

We can broadly identify two distinct dynamical regimes.
Given a typical energy scale $\omega$, for evolution times such that $\omega\tau\to\infty$, any continuous ramp is sufficient to achieve the target state, and the evolving state follows the instantaneous ground state of the system, as a consequence of the adiabatic theorem~\cite{Messiah}.
On the other hand, for very short evolution times $\omega\tau\ll 1$, the evolution is far from adiabatic.
In such regime, \emph{quantum speed limits}~\cite{Mandelstam:91,Giovannetti:03,Caneva:09,Taddei:13,Deffner:13,Deffner:17,Deffner:17r,Campbell:17,Deffner:20} provide fundamental bounds on the minimum evolution time $\tau$ required
to evolve between two states under a given time-independent dynamics.
 Such time is lower-bounded by a quantity proportional to the Bures angle between initial and final states, and inversely proportional of either the variance or the average energy along the trajectory.
It is worth stressing here that such quantum speed limits do not provide any information on  \emph{the optimal dynamics implementing the target transition}, but rather give an estimate of the evolution time \emph{for a given dynamics}.
The task of finding the optimal Hamiltonian achieving a given evolution is a more difficult problem, sometimes referred to as the \emph{quantum brachistochrone problem}~\cite{carlini2006timeoptimal} or \emph{minimum control time}~\cite{Poggi:19}.

The notion of control at the quantum speed limit has attracted considerable attention~\cite{Caneva:11,Hegerfeldt:13,Poggi:19}. In particular, it has been observed that the minimal evolution time to generate a ground state scales as $\tau^*\propto \Delta_{\min}^{-1}$ where $\Delta_{\min}$ denotes the minimum energy gap of the Hamiltonian during the evolution~\cite{Caneva:11a}.
%
%
This is particularly interesting for the LMG model, where $\Delta_{\min}$ occurs at the QPT and vanishes as $\Delta_{\min}\propto N^{-z}$ with $N$ the size of the system and $z=1/3$ the dynamical critical exponent~\cite{Dusuel:04}. 
However, we will provide examples in which this does not hold, and the minimal evolution time $\tau^*$ scales slower than $\Delta_{\min}^{-1}$, namely, $\tau^* \Delta_{\min}\propto N^{-\alpha}$ with $\alpha>0$ a scaling exponent.

\section{Optimal control}\label{sec:OC}
To find an optimal time-dependent protocol we define the cost function ${\cal F}_{\bs X}$ as the state fidelity between output and target state for a given protocol parameterisation $g_{\bs X}$
\begin{align}\label{eq:fcost}
{\cal F}_{\bs X}\equiv {\cal F}[g_{\bs X}]= \left\lvert\braket{\psi_{\bs X}(\tau)}{\phi_0(g_1)}\right|^2
\end{align}
where
\begin{equation}
	\ket{\psi_{\bs X}(\tau)}=U^{(\bs X)}_{\tau,0}\ket{\phi_0(g_0)}
	\equiv
	\mathcal{T} e^{-i\int_{0}^\tau  H\circ g_{\bs X}} \ket{\phi_0(g_0)}
\end{equation}
is the output state corresponding to a dynamics with pulse shape $g_{\bs X}$ and ${\cal F}_ {\bs X}=\left\lvert\braket{\psi_{\bs X}(\tau)}{\phi_0(g_1)}\right|^2$ is the state fidelity.
Here, $\ket{\phi_0(g)}$ is the ground state of $H(g)$, so that $\ket{\phi_0(g_0)}$ and $\ket{\phi_0(g_1)}$ are initial and target states, respectively, while ${\cal T}$ is the Dyson time-ordering operator.
Numerical optimization is used to maximise ${\cal F}_{\bs X}$ with respect to $\bs X$.
The different methods put forward to achieve this goal differ in how the function $g_{\bs X}$ is parameterised, that is, on the choice of \emph{ansatz} being considered.
Common choices include CRAB~\cite{Doria:11,Caneva:11}, local adiabatic ramps~\cite{Roland:02,Richerme:13} and bang-bang protocols~\cite{Balasubramanian:18,Cohn:18}.
Here we focus on bang-bang and in particular \emph{double}-bang protocols, benchmarking our results against those obtained via CRAB.

We also constrain the magnitude of the interaction $g_{\bs X}(t)$, imposing $\abs{g_{\bs X}(t)}\le g_{\max}$ for all $t$.
This ensures that the optimized protocols only require finite energy to be implemented, and ensures the existence of a maximum, i.e. non-zero, evolution time. We refer to App.~\ref{app:implementation_details} for the details about the employed optimization procedures. 

\subsection{Bang-bang protocols}\label{subsec:bang-bang}

\begin{figure}[b]
    \centering\hspace{-20pt}
    \includegraphics[width=\linewidth]{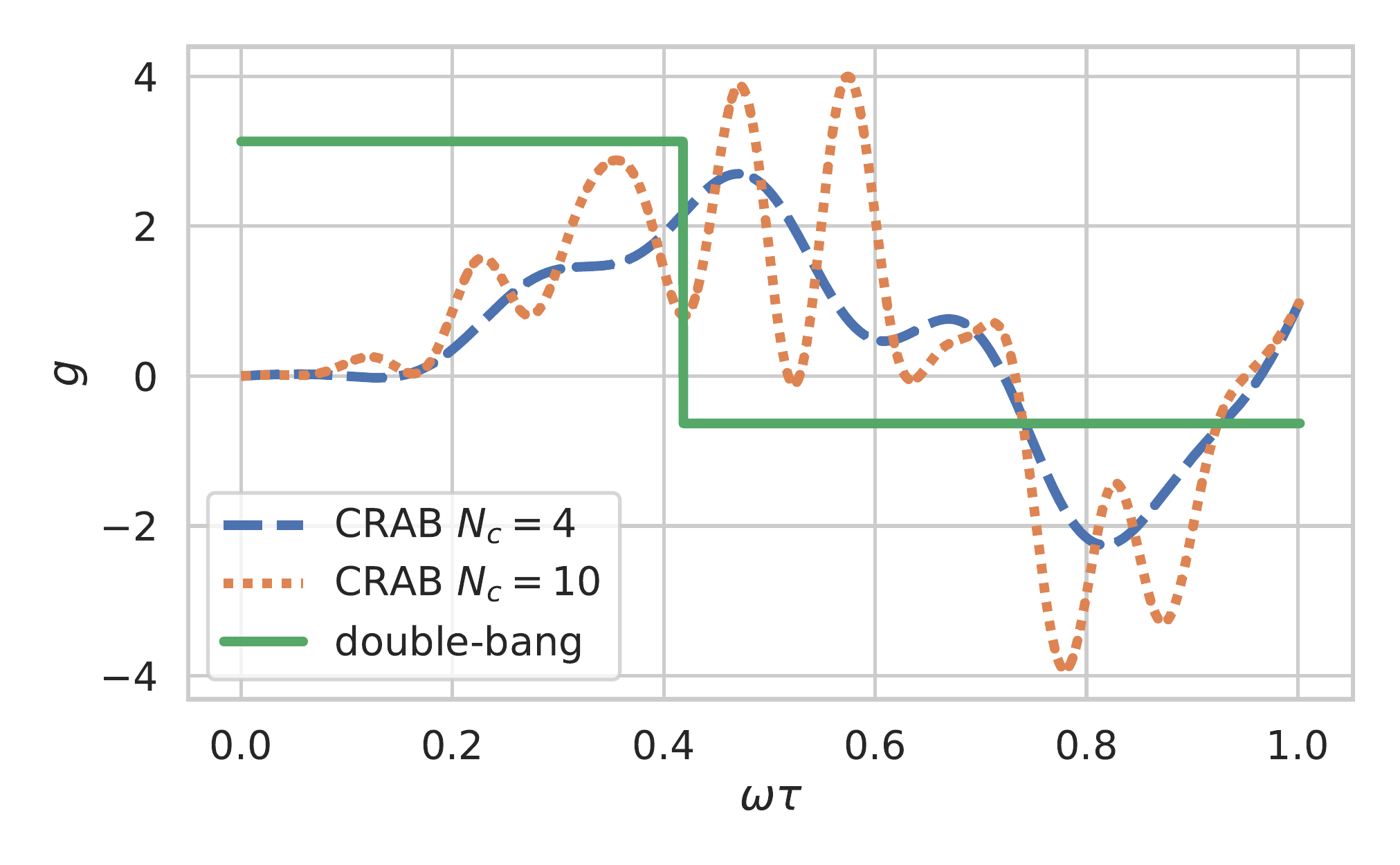}
    \caption{
    Comparison between optimal protocols for the generation of the critical ground state of an LMG model with $N=20$ spins as obtained using double-bang (green solid line) and CRAB with $N_c=4$ (dashed line) and $N_c=10$ (dotted line) frequencies. The specific values of the parameters used in these simulations, as well as the associated data can be found in Ref.~\cite{data}.}
    \label{fig:doublebang_vs_crab_protocols}
\end{figure}

\emph{Bang-bang protocols} with $\ell$ \emph{bangs} involve a piece-wise constant function of the form
\begin{equation}
\label{eq:BBgeneral}
	g^{\bs X,(t_1,t_{\ell+1})}_{\rm DB}(t) = \sum_{i=1}^\ell g_i \chi_{[t_i,t_{i+1}]}(t),
\end{equation}
where $\chi_I(t)=1$ for $t\in I$ and $\chi_I(t)=0$ otherwise.
Here, $t_1=0$ and $t_{\ell+1}=\tau$ are the fixed initial and final evolution times, respectively and $\bs X\equiv(g_1,...,g_\ell, t_2,...,t_\ell)$ are the $2\ell-1$ optimization parameters (with the added constraints $t_{i-1}\le t_i\le t_{i+1}$).
Note how bang-bang protocols involving $\ell+1$ bangs include as a subset bang-bang protocols with $\ell$ bangs.
In particular, the \emph{double}-bang protocols we will use have the form
\begin{equation}\label{eq:doublebang_protocol}
	g_{\rm DB}^{\bs X,(0,\tau)}(t) = \begin{cases}
		g_A, & 0\le t\le t_B, \\
		g_B, & t_B< t \le \tau,
	\end{cases}
\end{equation}
where $\bs X\equiv(g_A,g_B,t_B)$. When clear from the context, we will omit the explicit functional dependence of $g_{\rm DB}$ on its parameters, writing $g^{\bs X,(0,\tau)}_{\rm DB}\equiv g_{\rm DB}$.
In double-bang protocols, the control parameter $g(t)$ is thus instantaneously changed from $g_0$ to $g_A$ at the beginning of the protocol, then suddenly quenched to take value $g_B$ at some time $t_B$, and finally changed into $g_1$ at the end of the evolution~\footnote{Any experimental realization of these bang-bang protocols will inevitably require of a finite response time $\tau_r$, and thus instantaneous variations in $g(t)$ are just an idealization. However, provided $\tau_r\ll 1/\omega$ and $\tau_r\ll \tau$, with $\omega$ the typical energy scale of the system, such finite time response will not have any impact on the results presented here. }.
An example of a double-bang protocol is given in Fig.~\ref{fig:doublebang_vs_crab_protocols}. 
It is worth stressing that our use of the term \emph{bang-bang} differs from the way it is used in the context of NMR, where it refers to a technique to avoid environmental interactions~\cite{Viola:98}.

The piecewise-constant nature of the bang-bang protocols allows one to simplify the time-evolution operators, which can be written as
\begin{equation}
	U_{\tau,0}=e^{-i(\tau-t_B)H(g_B)}e^{-i t_B H(g_A)}.
\end{equation}
This makes simulating the associated dynamics computationally easier, compared to simulating the evolution of a state through generic time-dependent dynamics, as required for instance by  CRAB or Krotov protocols.

\subsection{Chopped-random basis quantum optimization (CRAB)}\label{subsec:CRAB}

In order to benchmark our results and highlight the advantages offered by the bang-bang protocols, we compare them with the results obtained via CRAB.
This method uses a time-dependent pulse shape written as a modulation of a linear ramp connecting initial and final parameter values. This variation is written in terms of trigonometric functions with randomly chosen frequencies.
More precisely, it uses the ansatz~\cite{Doria:11,Caneva:11,Rach:15,vanFrank:16}
\begin{equation}\scalebox{0.95}{$\displaystyle
g(t)=g_{\rm Lin}(t)\left[ 1+b(t)\sum_{n=1}^{N_c}\left(x_n\cos(\omega_n t)+y_n\sin(\omega_n t)\right)\right],
$}\end{equation}
where
\begin{itemize}
	\item
	$g_{\rm Lin}(t) \equiv g_0 +(g_1-g_0)t/\tau$
	is the linear ramp connecting $g_0$ and $g_1$ in a total time $\tau$.
	\item The integer $N_c$ is the total number of frequencies in the ansatz. Its value is set before the start of the evolution, together with the total evolution time $\tau$.
	\item The frequencies $\omega_n$ are uniformly sampled around the principal harmonics, $\omega_n=2\pi n\omega_0(1+\xi_n)$ with $\xi_n\in[-1/2,1/2]$ independent uniform random numbers.
	The use of random frequencies implies that the functional basis being used is not constrained to be orthogonal, a feature that was found to sometimes enhance the performance of the search algorithm~\cite{Doria:11,Caneva:11}.
	\item The function $b(t)$ is used to normalise the CRAB correction, ensuring $g(0)=g_0$ and $g(\tau)=g_1$. A possible choice for this is $b(t)=c t(t-\tau)$ for some constant $c>0$.
\end{itemize}
The optimization is run on the $2N_c$ parameters $\bs X\equiv (x_1,...,x_{N_c},y_1,...,y_{N_c})$.
Whereas $g_0,g_1,t_0,t_1$ are set by the problem, the values of  $\omega_n$ (equivalently, $\xi_n$) are chosen empirically (often randomly) before the evolution starts. The optimization algorithm is often further run for different sets of frequencies $\omega_n$, keeping only the best result.

Notice that while the most general formulation CRAB in principle encompasses a large class of parameterisations \cite{Caneva:11} which include bang-bang protocols as a special case, in this work we refer to the most common CRAB methods based on truncated random Fourier basis.

To compute the evolution of a state through a CRAB protocol we need to numerically simulate the dynamics through the time-dependent Hamiltonian. This is in general not as efficient as computing the evolution through piecewise-constant protocols.
Notice that the dynamics must be simulated a large number of times while looking for the optimal protocol, which builds up to a significant difference in computational times, as illustrated in the example addressed in Sec.~\ref{subsec:LZ}.


\begin{figure}[b]
    \centering
    \includegraphics[width=\linewidth]{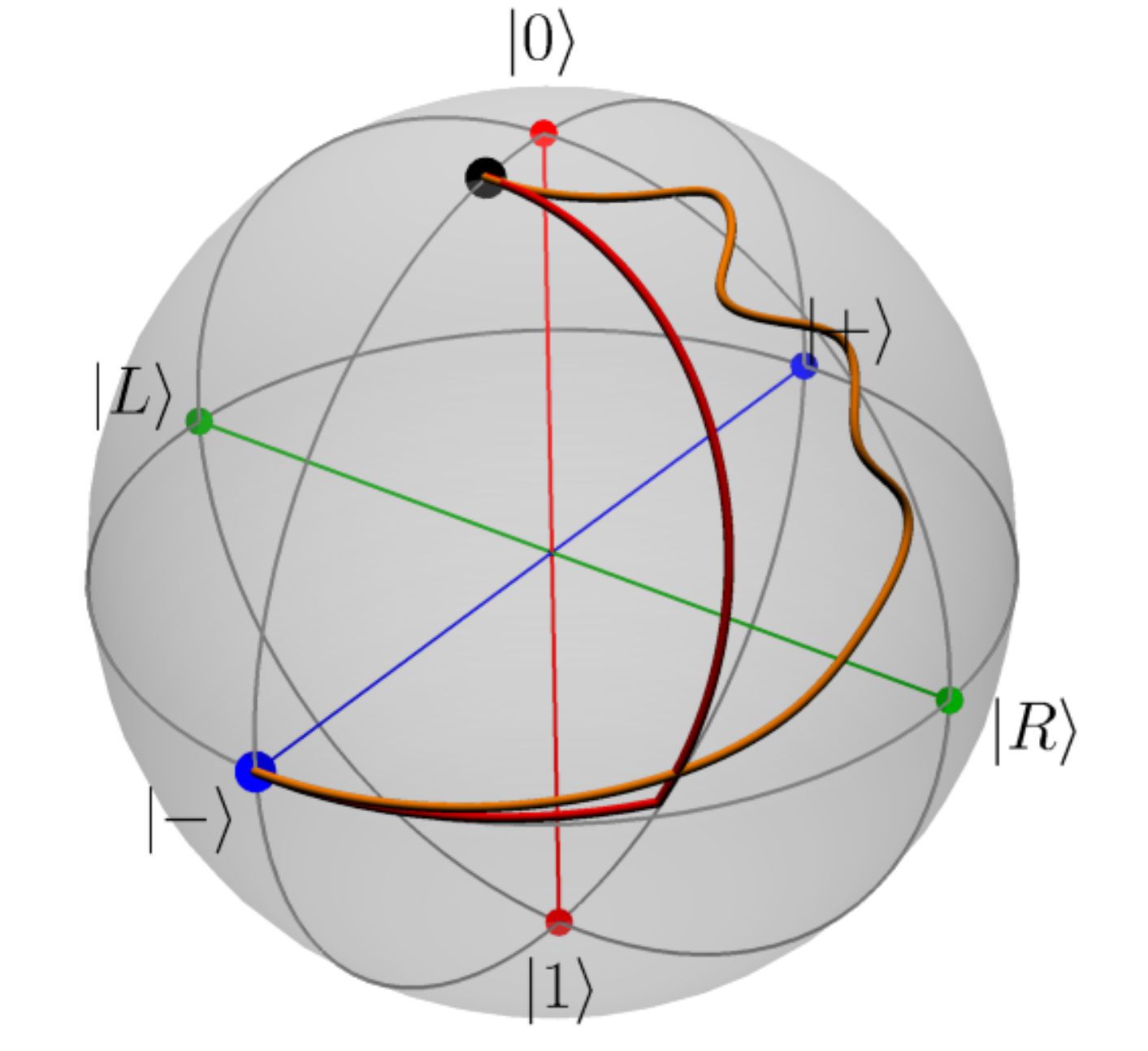}
    \caption{Representation of the dynamics corresponding to double-bang (red) and CRAB (orange) protocols, optimized to transport the ground state of a LZ model from $H=\omega \sigma_x+g_0\omega\sigma_z$ with $g_0=-5$ to $H=\omega\sigma_x$ with $g_1=0$. Note that $\sigma_x\ket{\pm}=\pm \ket{\pm}$, while $\ket{1}$ and $\ket{0}$ ($\ket{R}$ and $\ket{L}$) are the eigenstates of $\sigma_z$ ($\sigma_y$) with eigenvalue +1 and -1, respectively. The total evolution time is $\omega \tau=1$, and the CRAB protocol shown has $N_c=4$ frequencies. For such evolution time, both protocols reach the target state up to numerical precision.}
    \label{fig:lz_dbVScrab_onBlochSphere}
\end{figure}

\section{Applications}\label{sec:applications}

We here discuss the effectiveness of bang-bang protocols to generate the ground state at the critical point of LZ and LMG models, comparing them in particular with the results achieved using CRAB protocols. In~\cite{data} we make available the data as well as all the corresponding parameter values employed to generate the results.

\subsection{Landau-Zener problem}\label{subsec:LZ}

The LZ model describes a spin-${1}/{2}$ particle in a time-dependent magnetic field. The corresponding Hamiltonian reads
\begin{align}
H_{\rm LZ}(g)=\omega\sigma_x+g\omega\sigma_z = H_0 + g H_1,
\end{align}
with $H_0\equiv\omega \sigma_x, H_1\equiv\omega\sigma_z$, and $\sigma_k$ the $k$-Pauli matrix ($k=x,y,z$).
Without loss of generality, we set the initial state to be the ground state of $H_{\rm LZ}(-5)$,
and use as the ground state of $H_{\rm LZ}(0)$, that is the ground state at the avoided crossing, as a target.
The initial Hamiltonian $H_{\rm LZ}(-5)$ is an approximation of the asymptotic one $H_{\rm LZ}(-\infty)\sim -\sigma_z$. This approximation is sensible in this context, as the state fidelity between the ground states of $H_{\rm LZ}(-5)$ and $H_{\rm LZ}(-\infty)$ is $\sim0.99$.

We optimize over double-bang protocols for different evolution times. Our goal is to find simple protocols achieving the transition between initial and target ground states in the shortest possible time.
We thus scan different values of the evolution times $\tau$, optimizing the protocol for each chosen value. As shown in~\cite{Larocca:18}, depending on the imposed constraints and the time $\tau$ the control landscape shows a rich structure. 
We test both the bang-bang and CRAB protocols with different numerical optimization algorithms, and find that the double-bang protocols achieve better results in shorter $\tau$ times, while requiring significantly less computational time.
Studying the optimal protocols at several different times allows us to pinpoint the minimum value of $\tau$ required to reach the target state with our protocol, with a given fidelity.
We show in Fig.~\ref{fig:doublebang_vs_crab_protocols} examples of such optimized bang-bang and CRAB protocols (based on $N_c=4$ and $10$ frequencies). The first point to appreciate is the difference in the number of parameters that need to be optimized: while double-bang requires the management of only $3$ parameters, CRAB with $N_c=10$ frequencies needs $20$ coefficients (in addition to the frequencies in the optimization). This can be quite demanding for numerical optimization toolboxes, with differences in optimization times going from the order of hours for CRAB to seconds or minutes for double-bang. A second point of notice is that, intuitively, the search space grows with $N_c$, thus allowing CRAB to effectively encompass double-bang protocols. However, this would also make the associated optimization task demanding enough to be practically unfeasible. 

\begin{figure}
    \centering\hspace{-20pt}%
    \includegraphics[width=\linewidth]{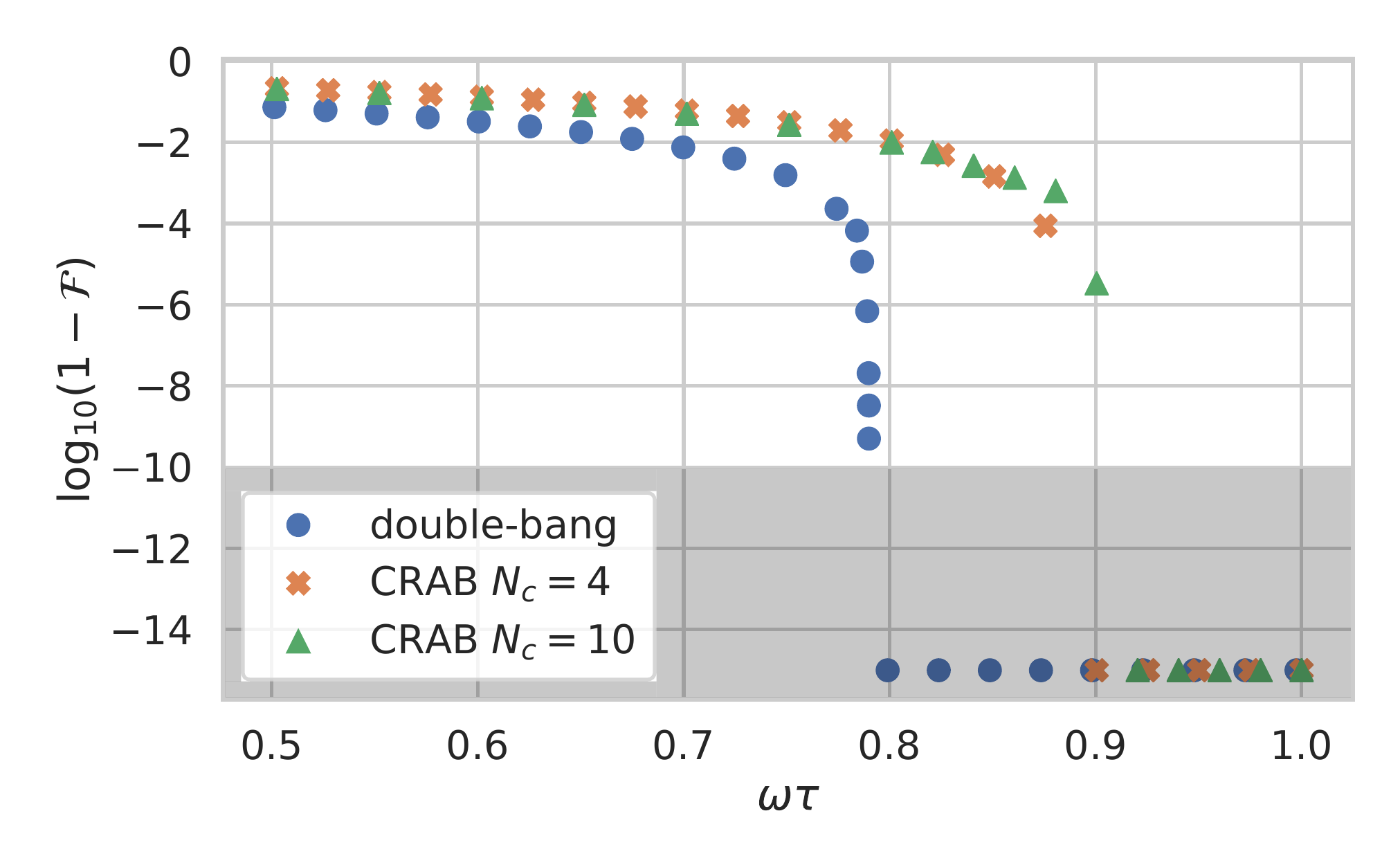}
    \caption{
    	Optimization results generating the ground state at the avoided crossing of an LZ model~\cite{data}.
    	We given the optimized fidelity $\mathcal F$ when using both double-bang and CRAB protocols for different total evolution times $\omega\tau$.
    	In each optimization, we constrain the available energy imposing $|g(t)|\le10 $ $\forall t$.
        Each point gives the fidelity obtained optimizing a double-bang (blue circles) or CRAB (orange crosses and green triangles) protocol to evolve the ground state of $H_{\on{LZ}}(-5)$ to the ground state of $H_{\on{LZ}}(0)$.
        The shaded region marks results for which the numerical precision starts being an important factor, and additional care must be taken to maintain the required level of accuracy while simulating the state.
        All the points shown in the figure correspond to
        $\mathcal F>1-10^{-14}$.
        Optimizations with up to $N_c=10$ CRAB frequencies do not provide a significant improvement in the fidelity.
    }
    \label{fig:lz_critical_crabAndDB}
\end{figure}

In Fig.~\ref{fig:lz_dbVScrab_onBlochSphere} we give a representation of the state evolution in the Bloch sphere under the protocols addressed here, while in Fig.~\ref{fig:lz_critical_crabAndDB} we report the fidelities obtained optimizing double-bang and CRAB protocols to achieve the ground state at the avoided crossing.
We find that double-bang, despite its simplicity, realises the target transition with good fidelity faster than CRAB, achieving fidelities ${\cal F}> 1-10^{-10}$ in time $\omega\tau^*\approx 0.8$.
whereas, CRAB requires $\omega\tau^*\approx0.9$ to reach similar fidelities.
We find that increasing the number of frequencies $N_c$ in CRAB does not bring about significant improvements, while making the optimization considerably more  computationally demanding.

To test further the minimum control time, we also performed the optimization with different protocols.
In particular, we tested a variation of CRAB in which initial and final values of the protocol are also included in the optimization, as well as \emph{triple-bang} protocols.
We find that both such approaches achieve ${\cal F}>1-10^{-10}$ at a shorter time $\omega\tau^*\approx 0.76$. This suggests that the sub-optimality of CRAB for this particular case might be partly due to the fixed initial and final parameter values and the inherent analyticity of the ansatz. 

\subsection{Lipkin-Meshkov-Glick model}\label{subsec:LMG}

The LMG model~\cite{Lipkin:65}, originally introduced in the context of nuclear physics, describes a fully long-range interaction of $N$ spin-${1}/{2}$ subjected to a transverse magnetic field.
Thanks to its experimental realisation with cold atoms~\cite{Zibold:10} and trapped ions~\cite{Jurcevic:17}, the model has gained renewed
attention~\cite{Titum:19,Louw:20,Fogarty:20,Puebla:20,Bao:20}, and has served as a test bed to study several aspects of quantum critical systems~\cite{Caneva:08,Yuan:12,Acevedo:14,Puebla:15,Campbell:16,Kopylov:17,Defenu:18}.
The model is described by the Hamiltonian
\begin{equation}
	H_{\rm LMG}(g)=\omega S_z-g\frac{\omega}{N}S_x^2
\end{equation}
with $S_{k}=\frac{1}{2}\sum_i \sigma_k^i$ the $k=x,y,z$ collective angular momenta operators. 
The model exhibits a second order mean-field QPT at a critical value $g_c=1$~\cite{Dusuel:04,Leyvraz:05,Vidal:07,Ribeiro:07,Ribeiro:08}
and belongs to the same universality class of the quantum Rabi~\cite{Hwang:15,Puebla:17} and the Dicke~\cite{Dicke:54} models. 

\begin{figure}
    \centering
    \includegraphics[width=\linewidth]{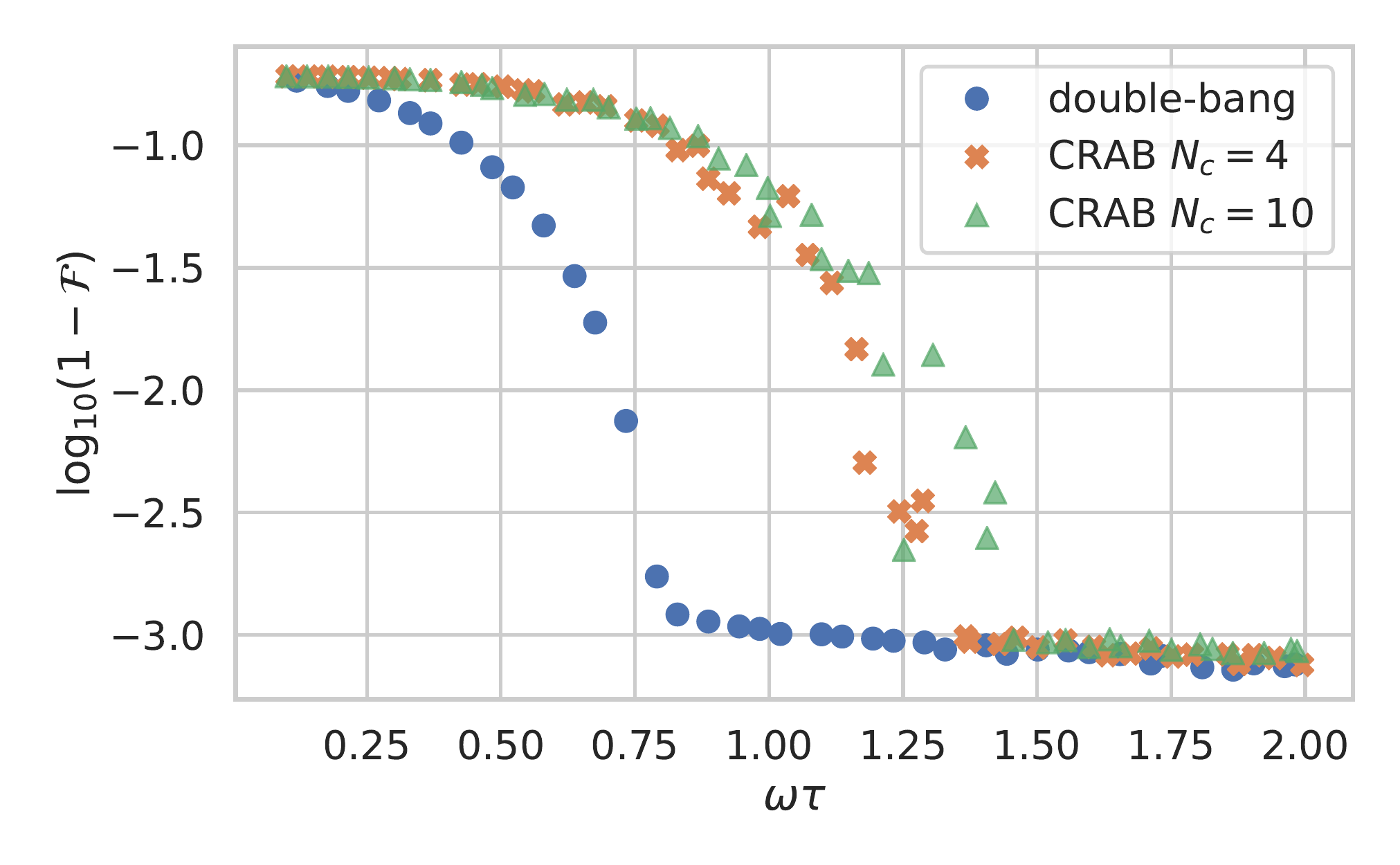}
    \caption{Results for the fidelity associated with the preparation of the critical ground state of the LMG for $N=50$ spins using a double-bang (blue circles) and CRAB protocols based on $N_c=4$ (orange crosses)  and $N_c=10$ (green triangles) frequencies. See main text and Ref.~\cite{data} for further details.}
    \label{fig:lmg_50spins_critical}
\end{figure}

\begin{figure}[b]
    \centering
    \includegraphics[width=\linewidth]{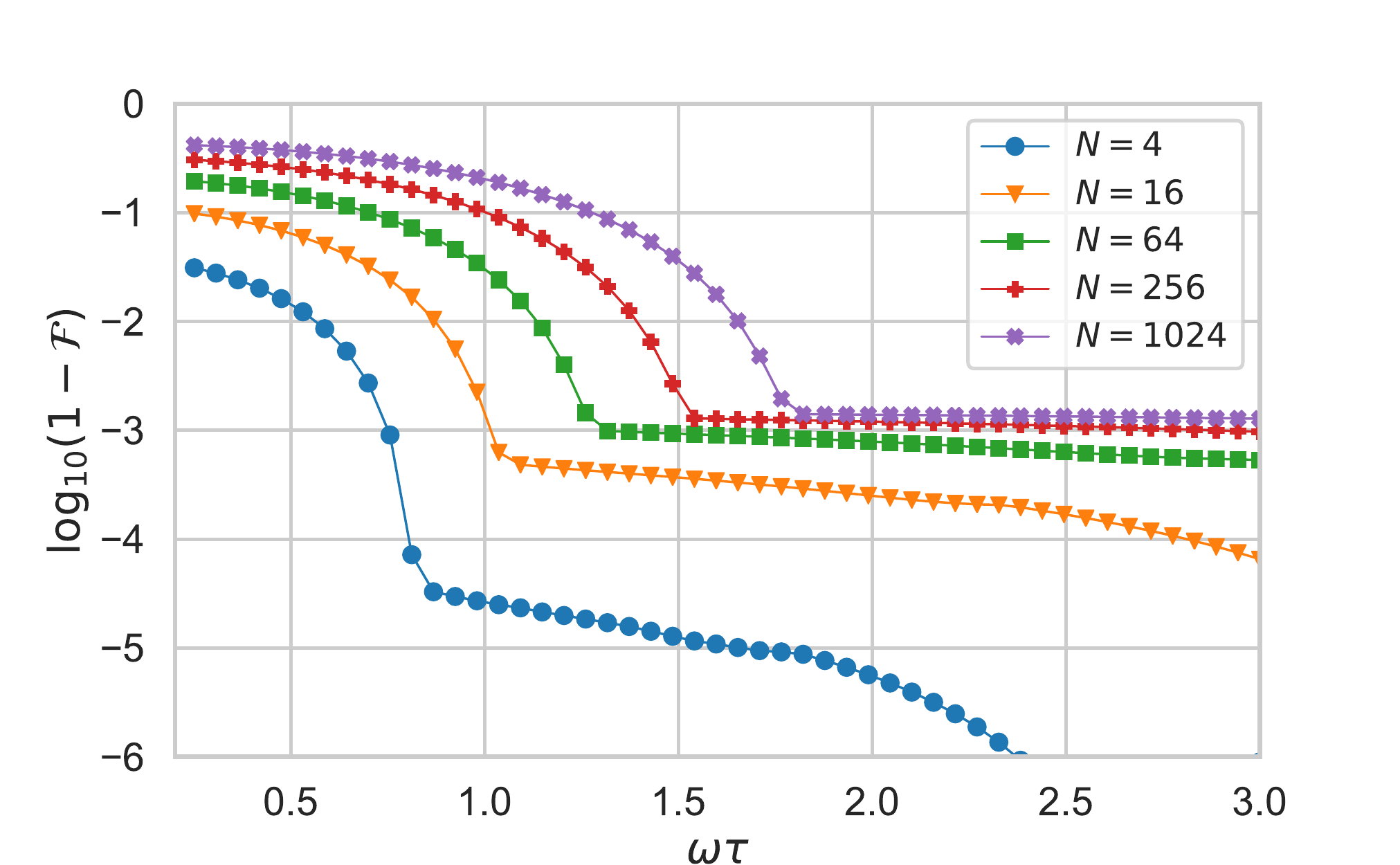}\hspace{-10pt}
    \caption{Fidelity as a function of the double-bang protocol duration $\omega\tau$ for a LMG model with $N=4, 16, 64, 256, 1024$ spins. See main text and Ref.~\cite{data} for further details.}
    \label{fig:lmg_doublebang_inf_vs_Nspins}
\end{figure}

\begin{figure}[tb]
    \centering
    \hspace{-20pt}
    \includegraphics[width=1.\linewidth,angle=-0]{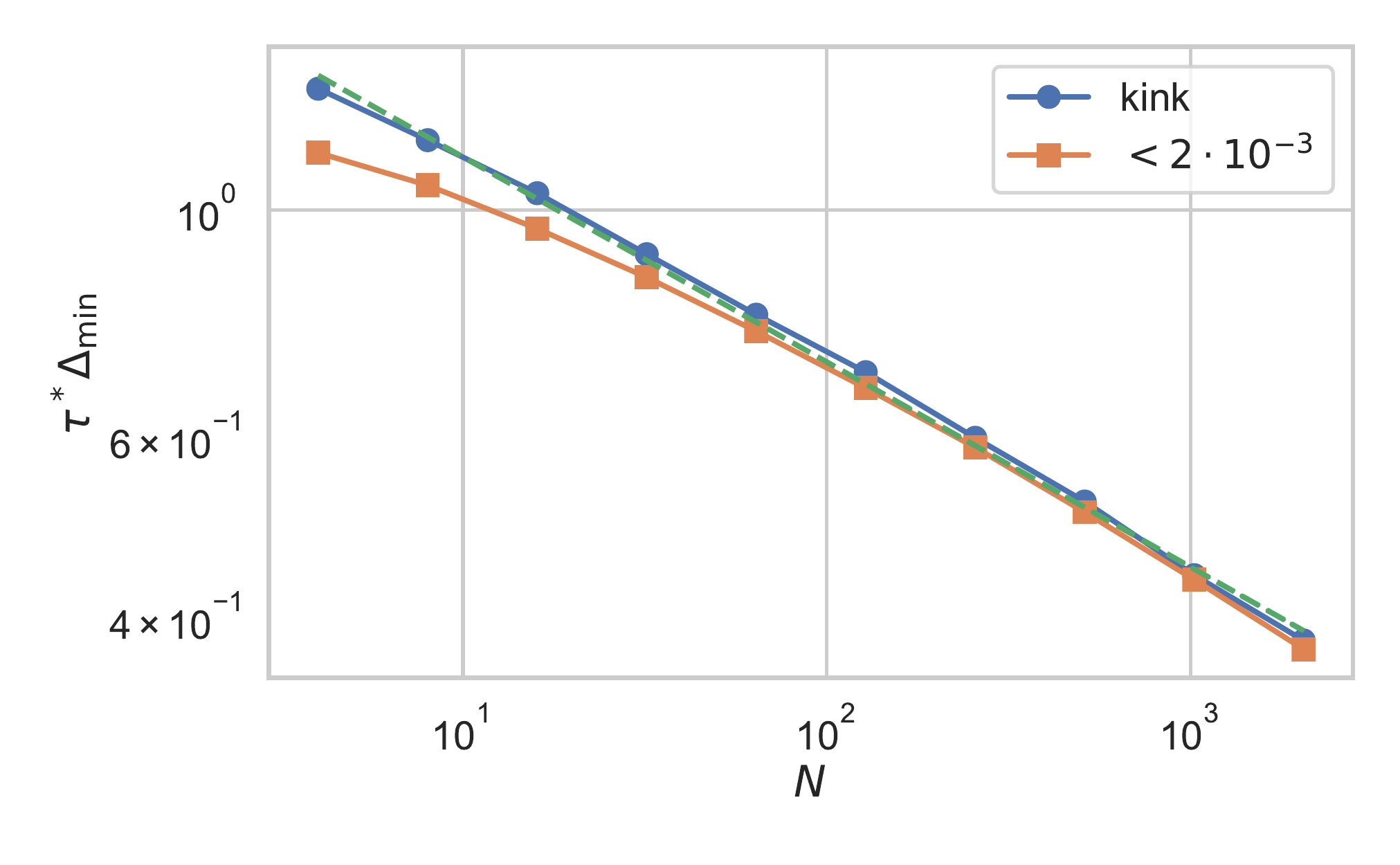}
    \caption{The scaling of $\tau^* \Delta_{\on{min}}$ versus the number of spins $N$, where $\Delta_{\on{min}}$ is the energy gap at the QPT and $\tau^*$ corresponds either to the location of the kink in ${\cal F}$, or when ${\cal F}>0.998$. The dashed line displays the fitted scaling law $\tau^*\Delta_{\min} \propto N^{-\alpha}$ with the exponent $\alpha=0.21(1)$ obtained for both quantities. We considered systems sizes up to $N=2048$ spins. See main text and Ref.~\cite{data} for further details.}
    \label{fig:lmg_scaling_kink_vs_N}
\end{figure}

\begin{figure}[b]
    \centering
    \includegraphics[width=\linewidth]{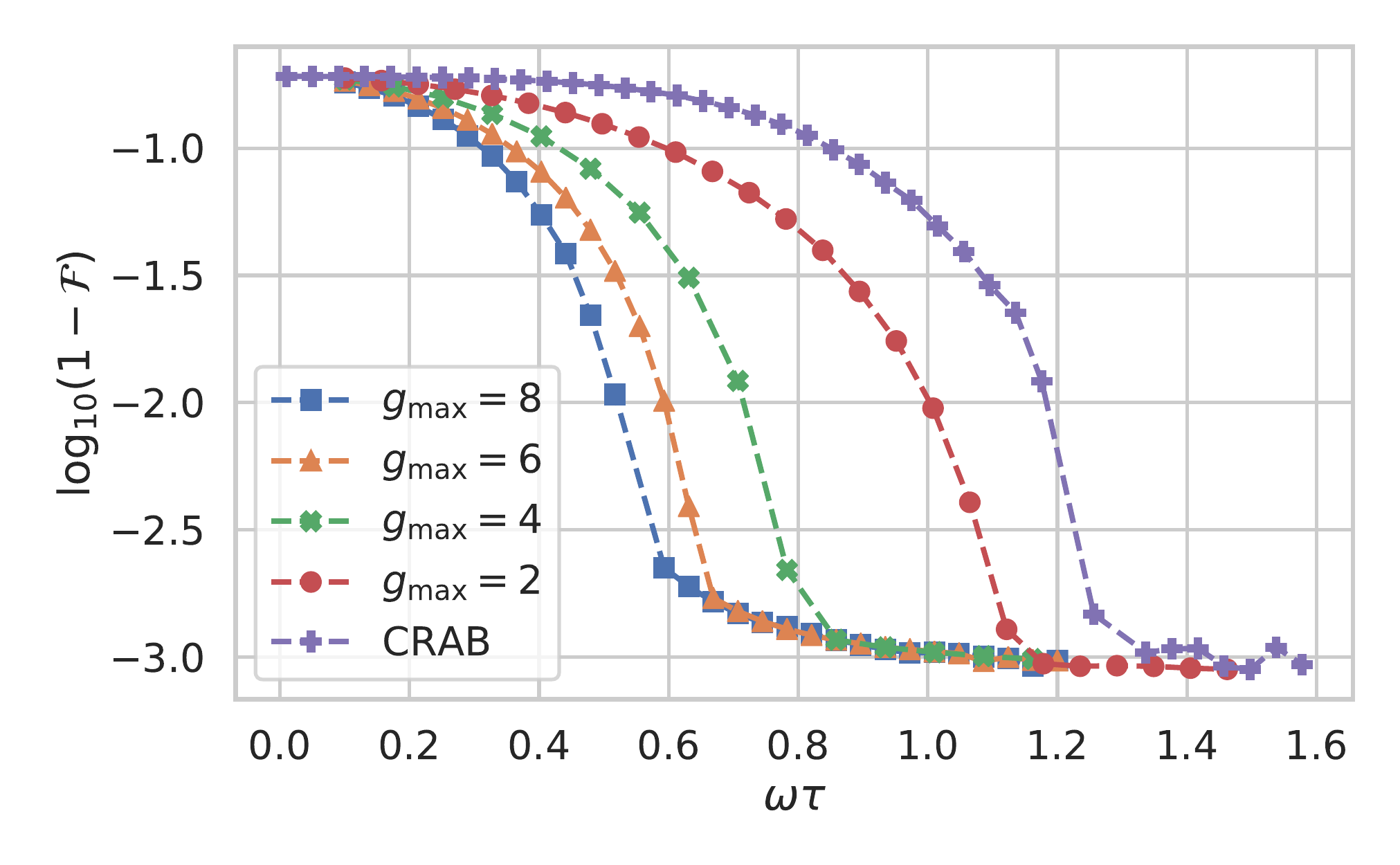}
    \caption{
		Optimizations of the double-bang protocol to achieve the critical ground state of the LMG model with $50$ spins, with different constraints applied to the total energy: $\abs{g(t)}\le g_{\max}$ with $g_{\max}=2,4,6,8$.
		For comparison, we also report the results achieved by CRAB with $4$ frequencies and bound $g_{\max}=8$. See main text and Ref.~\cite{data} for further details.
    }
    \label{fig:lmg_50spins_critical_diffBounds_withCrab_fixed}
\end{figure}

We focus on the task of driving the ground state of $H_{\rm LMG}(g_0=0)$ to the ground state at the critical point, $H_{\rm LMG}(g_c=1)$. See also Ref.~\cite{Ho:19} for a similar task using a variational quantum-classical simulation~\cite{Ho:19b}.  
As shown in Fig.~\ref{fig:lmg_50spins_critical}, in line with the results achieved for the LZ model, the double-bang protocols achieve the target transition with high fidelity faster than CRAB, and with scaling behavior better than the expected speed-limit scaling $\tau^*\sim \Delta_{\rm min}^{-1}$.
More precisely we find, with double-bang protocols, fidelities $\mathcal F\gtrsim 0.999$ for very short evolution times $\omega\tau\sim 1$. While $\mathcal{F}\gtrsim 0.99$ for $\omega\tau=0.75$ with a  double-bang protocol, CRAB with $N_c=10$ frequencies only achieves $\mathcal F\lesssim 0.9$ for the same $\tau$. Appendix~\ref{app:saturated_db} reports further details and results of the performance corresponding to the use of double-bang protocols.

Increasing the system size leads to a closing of the energy gap at the critical point according to $\Delta_{\rm min}\sim N^{-z}$ with $z=1/3$~\cite{Dusuel:04}. Hence, larger systems exhibit a smaller gap, which translates into longer evolution times to faithfully prepare the ground state at $g_c$. In Fig.~\ref{fig:lmg_doublebang_inf_vs_Nspins} we plot the results upon optimizing double-bang protocols for different system sizes. Without loss of generality, we choose a constraint $|g(t)|\leq g_{\rm max}=1.7$.
As argued in Ref.~\cite{Caneva:11}, an optimized protocol will only be able to find $\mathcal F\approx 1$ for protocols of duration $\tau\geq \tau^*\propto \Delta_{\rm min}^{-1}$. Since the energy gap scales as $N^{-z}$, it follows that $\tau^*\sim N^{z}$. Hence, $\tau^* \Delta_{\rm min}=O(1)$ should remain constant when the protocol operates at the quantum speed limit.
We find that the double-bang approach allows to prepare critical ground states with fidelities $\mathcal F\gtrsim  0.999$ in a time $\tau^* \Delta_{\rm min}\propto N^{-\alpha}$ with $\alpha>0$. We obtain an estimate of $\tau^*$ in two different ways: first, as the time at which the fidelity surpasses $\mathcal F=0.998$ and, second, as the time at which the kink displayed in Fig.~\ref{fig:lmg_doublebang_inf_vs_Nspins} in the fidelity is reached. Both criteria for $\tau^*$ lead to the same scaling, as shown in Fig.~\ref{fig:lmg_scaling_kink_vs_N} where we find $\tau^* \Delta_{\rm min}\propto N^{-\alpha}$ with $\alpha=0.21(1)$.

We also study the dependence of the minimal evolution time on the energy constraints imposed on the protocol.
As shown in Fig.~\ref{fig:lmg_50spins_critical_diffBounds_withCrab_fixed}, increasing the allowed energy decreases the minimum control time. Another interesting observed phenomenon is the existence of a \emph{threshold}, at around $\mathcal{F}\sim 0.999$, above which it is harder to push the fidelity. We find that the maximum fidelity, for both bang-bang and CRAB protocols, increases rapidly at first, but then hits a threshold, at which the increase is very slow with $\tau$.
Moreover, this threshold seems to be unaffected by the allowed energy, suggesting that it cannot be avoided by simply pumping more energy into the system, being instead related to the constraints inherent to the model under consideration. This same behavior can be seen also in Figs.~\ref{fig:lmg_50spins_critical} and \ref{fig:lmg_doublebang_inf_vs_Nspins}.

Our findings suggest the optimality of double-bang protocols for this task.
Even allowing for more complex protocols, we never find better fidelities than those achieved using the simple double-bang.
More precisely, we analyzed bang-bang protocols involving three and four \emph{bangs} [cf. Eq.~\eqref{eq:BBgeneral}],
finding no improvement with respect to the performance of double-bang protocols.
Indeed, it appears that the optimal protocols use only two distinct values of the parameter (as opposed to the three values allowed for by triple-bang protocols). This strongly suggests the optimality of a double bang for this task. As in the LZ case, this hints at a possible explanation of the sub-optimality of CRAB, which is constrained to use fixed initial and final values of $g(t)$. Optimal paths that involve a sudden quench at the beginning and/or end of the protocol are hardly attainable with a continuous CRAB with a finite number of frequencies. 
As further evidence in this direction, we considered a variation of CRAB in which the endpoints $g_0,g_1$ are also optimized. Consistently with our conjecture, this improves the results, pushing the minimal time for $N=50$ spins to $\omega\tau^*\approx1$. As it can be seen in Fig.~\ref{fig:lmg_50spins_critical},  $\omega\tau^*$ obtained with optimized endpoints lies between the $\omega\tau^*$ achieved with double-bang and that of CRAB with fixed initial and final points.

\section{Conclusions}\label{sec:conclusions}

We have shown that simple double-bang protocols can be employed for a fast and faithful ground state preparation. In particular, we have explicitly addressed the paradigmatic LZ and LMG models, the latter to illustrate the possibility to reliably prepare a critical ground state. In these models, optimized double-bang protocols can perform better than well-established optimal control techniques, such as CRAB. Owing to their nature, these double-bang protocols are very well suited for optimization purposes, offering a large computational advantage with respect to other optimal control methods.

In the LMG model, double-bang protocols allow the preparation of the ground state at the critical point in a time that scales slower than the inverse of the energy gap at the QPT. Other quantum critical models can be investigated following similar routes. Although distinct optimal control techniques may reach similar results than those reported here under a double-bang scheme, the often large number of variables to be maximised makes these protocols very difficult to be optimized, thus hindering this key observation.

Our results motivate further theoretical studies in the realm of quantum speed limits in many-body systems. It is worth stressing that our double-bang protocol can be readily implemented in different experimental setups, allowing for the fast preparation of interesting quantum states, such as highly entangled states of a large number of ions~\cite{Cohn:18}.


\begin{figure*}
    \centering
    \begin{tikzpicture}
    	\node  at (0, 0)%
			{\includegraphics[width=\linewidth]{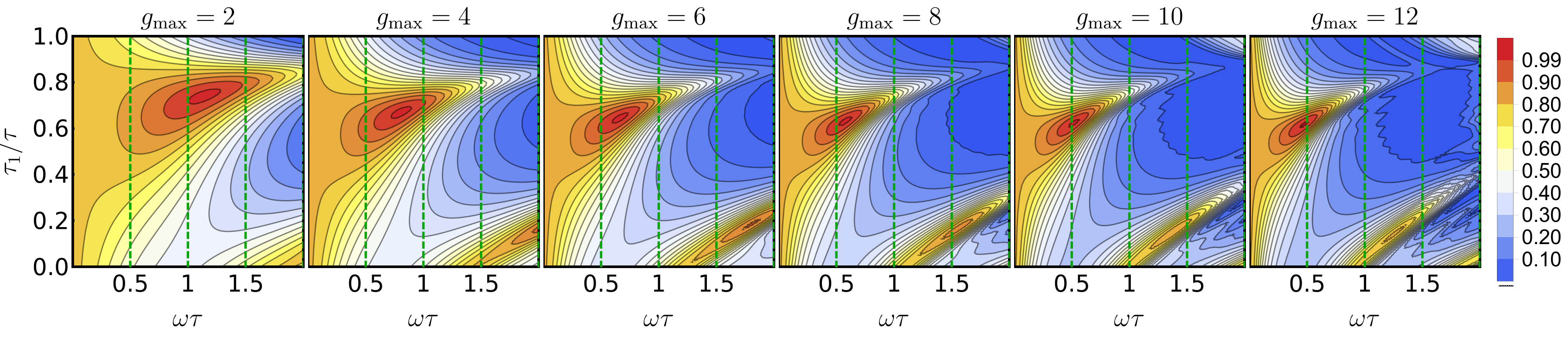}};
    	\node at (8.28, 1.77) {$\mathcal F$};
    \end{tikzpicture}
    \caption{
        Saturated double-bang protocol scans for $N=50$ spins.
        In each plot we report the fidelity corresponding to a saturated double-bang protocol, in which the interaction strength is $g_{\on{max}}$ for times ranging from $0$ to $\omega\tau_1$, and $-g_{\on{max}}$ between $\omega \tau_1$ and $\omega\tau$.
        All plots use the same color scale, with dark blue corresponding to values close to zero and bright red to fidelities $\mathcal F>0.99$.
        The dashed vertical green lines are only used to mark the values $\omega\tau=0.5$, $1.0$ and $1.5$.}
    \label{fig:lmg_satDoubleBangScans_N50}
\end{figure*}

\begin{acknowledgments}
The authors acknowledge helpful discussions with Tommaso Calarco, Steve Campbell and Simone Montangero, as well as the support by the SFI-DfE Investigator Programme (grant 15/IA/2864), the UK EPSRC (grant number EP/S02994X/1), the H2020 Collaborative Project TEQ (grant agreement 766900), the Leverhulme Trust Research Project Grant UltraQuTe (grant RGP-2018-266) and the Royal Society Wolfson Fellowship (RSWF/R3/183013). 
\end{acknowledgments}

\appendix
\section{Implementation details}
\label{app:implementation_details}

The optimizations reported here have been carried out in \textsc{Python}, using the algorithms provided by the \textsc{Scipy} scientific library~\cite{virtanen2020scipy}.
We used the \emph{Nelder-Mead}~\cite{nelder1965simplex} and \emph{Powell}~\cite{powell1964efficient} optimization methods, which were found to give the best performances.
Nelder-Mead, in its \emph{adaptive} variant~\cite{gao2010implementing}, is found to give better results when using the CRAB protocol, while Powell gives better results to optimize bang-bang protocols.

\begin{figure}
    \centering
    \begin{tikzpicture}
    	\node  at (0, 0)%
			{\includegraphics[width=0.85\linewidth]{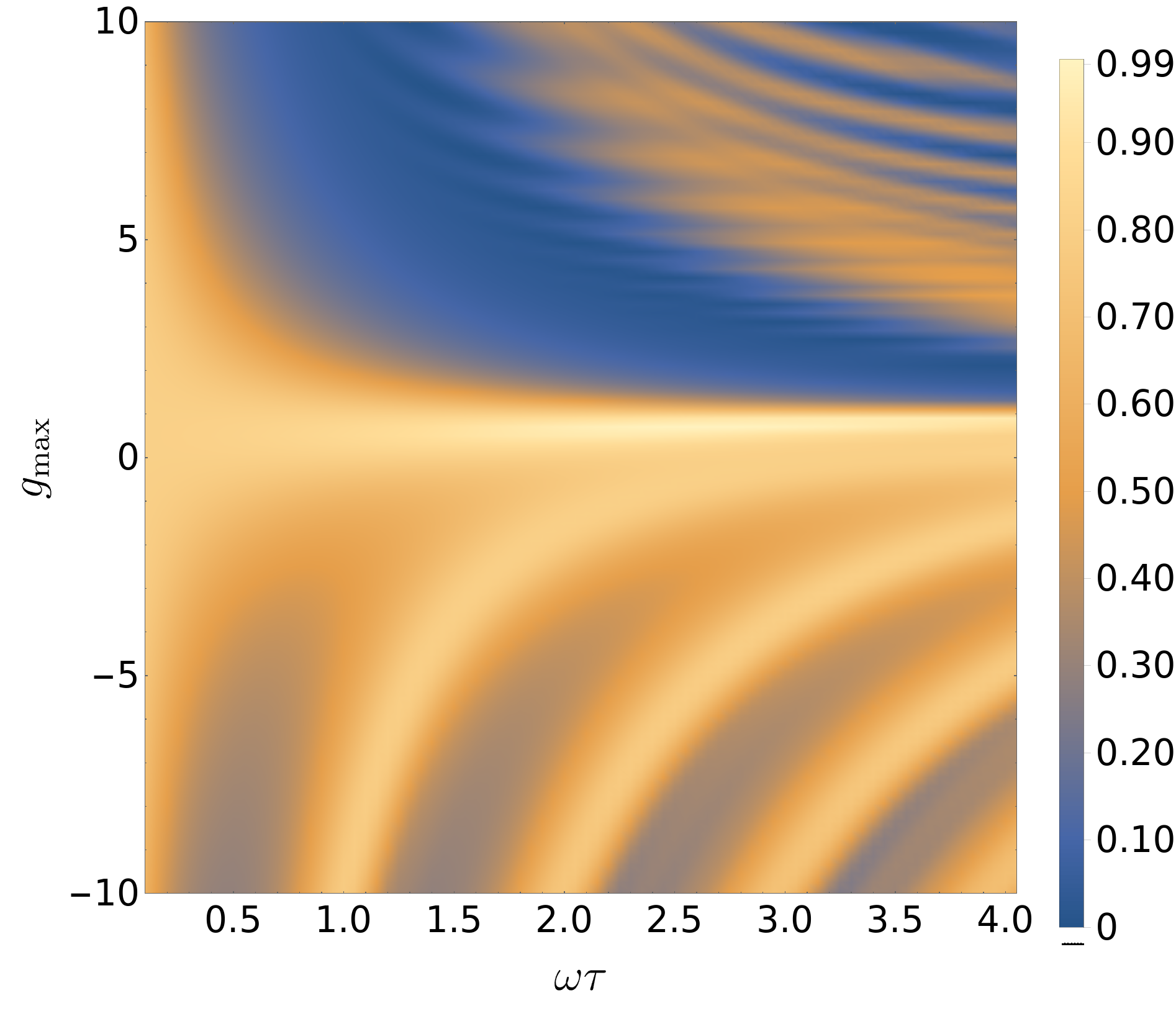}};
    	\node at (3.1, 3.1) {$\mathcal F$};
    \end{tikzpicture}
    \caption{
        Fidelity as a function of the evolution time, for the LMG model with $N=50$ spins, evolving following a constant protocol.
        For each total evolution time $\omega\tau$ and value of $g_{\rm max}$, we report the corresponding fidelity.
        The optimal value of the fidelity is achieved for all times at values of $g_{\rm max}$ between $0.5$ and $0.9$. Recall that the QPT takes place at $g=1$.}
    \label{fig:lmg_50spins_constantparameters_scan10}
\end{figure}

\section{Saturated-boundary double-bang protocols}
\label{app:saturated_db}

In the phase in which the optimal fidelity increases quickly, before the saturation point, the optimal double-bang protocols are found to be of the following form: $g(t)=g_{\rm max}$ for $t\in[0,\tau_1]$, for $\tau_1$ some threshold time, and $g(t)=-g_{\rm max}$ for $t\in[\tau_1,\tau]$, with $\tau$ the total evolution time.
We analyse this further in Fig.~\ref{fig:lmg_satDoubleBangScans_N50}, where for different energy constraints $g_{\rm max}$, we show the fidelities for the different possible \emph{saturated double-bang} protocols, by varying $\omega\tau$ and $\tau_1/\tau$ to explore the different possible shapes.
We find that the saturation threshold observed in Figs.~\ref{fig:lmg_50spins_critical}, \ref{fig:lmg_doublebang_inf_vs_Nspins} and \ref{fig:lmg_50spins_critical_diffBounds_withCrab_fixed} corresponds to a marked change in behavior of the fidelity. 
Although not explicitly shown, we analyse the scaling of the optimal time $\omega \tau^*$ as a function of the energy constraint $g_{\rm max}$, which is found to follow  $\omega\tau^*=1.819\cdot g_{\on{max}}^{-0.559}$, where the values are determined via a numerical fit. 
For completeness, we also give in Fig.~\ref{fig:lmg_50spins_constantparameters_scan10} the fidelities obtained using a \emph{constant} protocol based on the use of a $g_{\rm max}$ value. As expected, in this simple model it is not possible to exploit the available energy to speed up the transition, and fidelities $\mathcal{F}>0.99$ are only possible for small energies, and the times always larger than those obtainable using double-bang.

%

\end{document}